\documentstyle[12pt]{article}
\begin{document}
\title{"QUANTUM MECHANICAL BLACK HOLES: ISSUES AND RAMIFICATIONS"}
\author {B.G. Sidharth\\ Centre for Applicable Mathematics \& Computer Sciences\\
B.M. Birla Science Centre, Hyderabad 500 063 (India)\\
Dedicated to the memory of my parents}
\date{}
\maketitle
\footnotetext{Proceedings of the International Symposium on "Frontiers of
Fundamental Physics", Universities Press (in press).}
\begin{abstract}
The recent model of Quantum Mechanical Black Holes is discussed and its
implications for cosmology, particle structure, low dimensionality and
other issues are examined.
\end{abstract}
\section{The "Ganesha" Quantum Mechanical Black Hole Model}
It was shown in a previous communication$^1$ that a typical elementary
particle, the electron can be considered to be what was termed a Quantum
Mechanical Black Hole (or QMBH), made up of a relativistic fluid of subconstituents,
or "Ganeshas" described by the Kerr-Newman metric giving both its gravitational and
electromagnetic fields and also the anomalous gyromagnetic ratio$^2$. Alternatively the QMBH could
be described as a relativistic vortex in the hydrodynamical formulation. It
was pointed out that the QMBH or vortex could also be thought of as a
relativisitc rotating shell reminiscent of the earlier Dirac model$^3$.\\
However the horizon of the Kerr-Newman Black Hole becomes in this case
in an obvious notation, complex,
\begin{equation}
r_+ = \frac{GM}{c^2} + \imath b,b \equiv (\frac{G^2Q^2}{c^8} + a^2 -
\frac{G^2M^2}{c^4})^{1/2}\label{e1}
\end{equation}
In the Quantum Mechanical domain, (\ref{e1}) can be seen to be meaningful,
if we remember that , the position coordinate for a Dirac particle is
given by$^4$
\begin{equation}
x_\imath = (c^2p_\imath H^{-1} t + a_\imath)+\frac{\imath}{2}
c\hbar (\alpha_\imath - cp_\imath H^{-1})H^{-1},
\label{e2}
\end{equation}
where $a_\imath$ is an arbitrary constant and $c\alpha_\imath$ is the velocity operator
with eigen values $\pm c$. The real part in (\ref{e2}) is the usual position
while the imaginary part arises from Zitterbewegung. Interestingly, in both
(\ref{e1}) and (\ref{e2}), the imaginary part is of the order of $\frac{\hbar}
{mc}$, the Compton wavelength, and leads to an immediate identification of
these two equations. It was pointed out in ref.1 that our physical measurements
are really averaged over a width
of the order $\frac{\hbar}{mc}$. So also with time measurements.
That is in the case of the QMBH (Quantum Mechanical Black Hole), obtained by
identifying (\ref{e1}) and (\ref{e2}), the naked singularity is shielded
by a Quantum Mechanical averaging which leads to real eigen values.\\
Further,$^5$ at the Compton wavelength
it is the negative energy two spinor $\chi$ that dominates, which under reflection,
behaves like a psuedo-spinor.
Hence the operator $\frac{\partial}{\partial x^\mu}$ acting on $\chi$, a
density of weight $N = 1,$ shows up as the electromagnetic field with
discrete charge:
An electron can be associated with curvature and the double
connectivity of spin half is reconciled with geometrodynamics.
So we could treat the Quantum Mechanical Black Hole as a relativistic fluid
of subconstituents (or Ganeshas). In a linearized theory (cf.ref.2)
it was shown, (cf.ref.1), that not only do we recover the Quantum
Mechanical spin but also that we get, the well known empirical result,
\begin{equation}
\frac{e^2}{Gm^2} \sim 10^{40}\label{e3}
\end{equation}
The above model gives a rationale for the left handedness of the neutrino,
which can be treated as an electron with vanishing mass so that the
Compton wavelength becomes arbitrarily large. For such a particle, we
encounter in effect the region within the Compton wavelength with the pseudo
spinorial property discussed above, that is left handedness. Further, in the
absence of the spinorial Compton wavelength boundary, an anomalous Bosonic
behaviour results.\\
It may be remarked that the electron, the positron and its special case the neutrino
are the fundamental elementary particles which could be used to generate
the mass spectrum of elementary particles$^{6,7}$.
We observe that in this formulation the Compton wavelength emerges as a fundamental
length, and so also $\hbar/mc^2$ emerges as a fundamental unit of time
reminiscent of Caldirola's chronon formulation.\\
To see how QMBH can be formed, we observe that as
is well known there is a zero
point field (ZPF) and the energy of the fluctuations of the magnetic field in a region of length
$\lambda$ is given by$^{2}$ $(\vec E$ and $\vec B$ are electromagnetic field
strengths)
\begin{equation}
B^2 \sim \frac{\hbar c}{\lambda^4}\label{e4}
\end{equation}
If $\lambda$ as in the QMBH is taken to be the Compton wavelength
$\frac{\hbar}{mc}$, (\ref{e4}) gives us for the energy in this volume
of the order $\lambda^3$,\\
$$\mbox{Total\quad energy\quad of\quad QMBH\quad}\sim \frac{\hbar c}{\lambda} =
mc^2,$$
exactly as required. In other words the entire energy of the QMBH
of mass $m$ can be thought to have been generated by the fluctuations
alone (cf.ref.5).
There is a further justification for the above interpretation. Let us use in (\ref{e4})
the pion Compton wavelength as the cut off, because the pion
is considered to be a typical elementary particle.
Then  we can recover the pion mass,
$m_\pi$ and moreover,
\begin{equation}
Nm_\pi = M,\label{e5}
\end{equation}
where $N$ is the number of elementary
particles, typically pions, $N \sim 10^{80}$ and $M$ is the mass
of the universe,viz. $10^{56}gms$. This also avoids certain divergences
encountered in QED.\\
\section{Cosmological Implications}
The question that arises is, if we treat the entire universe as arising from
fluctuations, is this picture consistent with observation? This is so.
Equation (\ref{e5}) is the first of the correspondences.
We can next deduce using the ZPF spectral density, the relation (cf.ref.5),
$M \alpha R$ where $R$ is the radius of the universe. This is quite correct and infact poses a puzzle, as is
well known and it is to resolve this dependence that dark matter has
been postulated whereas in our formulation the correct mass radius
dependence has emerged quite naturally.
Other interesting and consistent  consequences are as follows (cf.ref.5):
\begin{equation}
\frac{GM}{c^2} = R, \sqrt{N} = \frac{2m_\pi c^2}{\hbar} .T\label{e6}
\end{equation}
where $T$ is the age of the universe $\approx 10^{17}secs$, and,
\begin{equation}
H = \frac{Gm_\pi^3 c}{\hbar^2},\label{e7}
\end{equation}
It is remarkable that equation (\ref{e7}) is known to be true from a
purely empirical standpoint is considered mysterious.
We can also deduce that,
\begin{equation}
\frac{d^2R}{dt^2} = H^2R \equiv \Lambda R\label{e8}
\end{equation}
That is, effectively there is a cosmic repulsion,
which is not only consistent but numerically agrees exactly with the limit on this constant
(cf.ref.2).
To proceed we observe that the fluctuation of $\sim \sqrt{N}$ in the
number of particles leads to (\ref{e3}),
$$N_1 = \frac{e^2}{Gm^2} \approx 10^{40},$$
with $N_1 = \sqrt{N},$ whence we get,
\begin{equation}
R = \sqrt{N}l, \frac{Gm}{lc^2} = \frac{1}{\sqrt{N}}, G \propto T^{-1},\label{e9}
\end{equation}
as also the fact that $\dot G/G \sim \frac{1}{T}$, in reasonably good agreement.\\
Further, from the above we can deduce that the charge $e$ is independant
of time or $N$.
Infact we can treat $m$ (or $l$), $c \mbox{and}\hbar$ as the only microphysical
constants and $N$ as the only cosmological parameter, given which all other
parameters follow.\\
We can now easily deduce from (\ref{e6}), (\ref{e7}) and (\ref{e8}), the following:
\begin{equation}
\rho \propto T^{-1}, \Lambda \propto T^{-2}, M = \frac{c^3}{GH}\label{e10}
\end{equation}
a relation which appears in the Friedman model of the expanding universe
(the critical closure density) and the Steady State model also with $k = 0,$
that is in a flat universe. This is not surprising because Friedman Cosmology
as here, is apart from isotropy, matter dominated. But in the Steady State
theory, $\rho$ remains constant while in our model it tends to zero from
(\ref{e10}).\\
In our model the equation (\ref{e6}) actually provides an
arrow of time, atleast at the cosmological scale, in terms of the particle
number $N$.
Secondly, more of the fluctuationally created particles appear for example
near Galactic centres, than in empty voids, reminiscent of the jets which are
observed. Next, the farther out into space we look, the greater is $G$ there.
This poses a correction on the masses estimated by using the same value
of $G$. Finally, the cosmic background radiation can be explained in terms of
fluctuations of Boltzmann's $H$ function (ref.5).
\section{The structure of particles}
Our next task is to exhibit how other particles can be built up from electrons
and neutrinos (and their anti particles). The question is, are the groupings
of particles as fundamental as the shapes of stellar constellations?
It was shown in reference 7 that we can consider the proton to be made
up of two positrons separated by a central electron at a distance $r = l$,
the Compton wavelength of the proton, from the other two particles.
It must be pointed out that this is borne out by deep inelastic scattering data.
We next observe that the $\pi^+$ and $\pi^-$ can be seen to be bound states
of a positron or an electron and a neutrino respectively:
It has been pointed out earlier that the neutrino has an infinite or very large
Compton wavelength only when it is not in a bound state. When bound the
neutrino's Compton wavelength is no longer infinite and it acquires
a small mass $m'$, and in the QMBH model a small charge $e'$ such that, by (\ref{e3})
$$\frac{e'^2}{m'^2} = \frac{e^2}{m_e^2} = G.10^{40}$$
Let us now consider a bound state of an electron and a neutrino with the
latter orbiting the former at a distance $r$ which is the Compton wavelength
of the resultant particle. Equating the electrostatic and centrifugal
forces we get,
$$r = \frac{e^2}{m_ec^2}$$
which is the Compton wavelength of the pion.
Thus the resulting particle in the $\pi^-$ and similarly the $\pi^+$ is
a bound state of a positron and a neutrino. The $\pi^\circ$ is a positronium
type of a bound state of an electron and a positron (cf.ref.1), on the lines
of two photons.
The decay modes of the three pions bear out the above scheme This apart when
a beam of $\pi^\circ$ (or photons) encounters powerful electric or magnetic
fields, we would expect the production of electrons and positrons.
The muon as above could be considered to be made up of a neutrino orbiting the $\pi^-$.
Remembering that the
$\pi^-$ itself is a composite particle, the muon Compton wavelength is $3/2$ times roughly the $\pi$ Compton wavelenth
in agreement with observation.
Again the decay mode of the muon particles is
in agreement with the suggested combinations. So also the other
particle masses emerge (cf.ref.6).\\
\section{Weak Interactions}
We saw that the neutrino exhibits an anomalous Bosonic behaviour.
If we use Bose statistics for neutrinos, it follows that,
\begin{equation}
\frac{m_\nu c^2}{k} \approx \sqrt{3}T\label{e11}
\end{equation}
At the present background temperature of about $2^\circ K$, this
gives a neutrino mass
\begin{equation}
10^{-9}m_e \le m_\nu \le 10^{-8}m_e\label{e12}
\end{equation}
where $m_e$ is the electron rest mass.
It is remarkable that (\ref{e11}) is exactly what is required to be deduced
theoretically to justify recent models of lepton conservation or in certain
unification schemes.
We now observe that the balance of the gravitational force
and the Fermi energy of these cold background neutrinos, gives$^8$,
$$\frac{GN_\nu m_\nu^2}{R} = \frac{N_\nu^{2/3}\hbar^2}{m_\nu R^2},$$
whence, $N_\nu \sim 10^{90}$\\
where $N_\nu$ is the number of neutrinos, which is correct.\\
If the new weak force is mediated by an intermediate particle of mass $M$ and
Compton wavelength $L$, we will get from the fluctuation of the particle
number $N$, on using (\ref{e12}),
\begin{equation}
g^2\sqrt{N_\nu} L^2 \approx m_\nu c^2 \sim 10^{-14},\label{e13}
\end{equation}
From (\ref{e13}), on using the value of $N_\nu$, we get,$g^2L^2 \sim 10^{-59}$\\
This agrees with experiment and the theory of massless particles the neutrino
specifically acquiring mass due to interaction$^9$, using the usual
value of $M \sim 100 Gev.$
Additionally there could be a long range force also, a "weak electromagnetism"
with coupling   $\bar g$. This time, in place of (\ref{e13}), we would have,
\begin{equation}
\frac{\bar g^2 \sqrt{N_\nu}}{R} \approx 10^{-8}m_\nu c^2\label{e14}
\end{equation}
Comparing (\ref{e14}) with a similar equation for the electron, we get,
$$\bar g^2/e^2 \sim 10^{-13}$$
so that the neutrino will appear with an electric charge a little less than
a millionth that of the electron.\\
\section{Other issues:}
1. It was shown in reference 1, that at the electron's Compton wavelength,
we encounter a QCD type potential. This can be approximated by a Gaussian
type potential, $\frac{A}{r} e^{-\mu^2 r^2},
\mu = \frac{mc^2}{\hbar},$ whence the
electron has a "size" of $10^{-21}cm$ in agreement with the results of
Dehmelt and co-workers.\\
2. If, as in
our model, spin half is fundamental, it is known that this leads to a space with minimum three
dimensions (ref.2). However if we consider electrons or more generally Fermions in an idealised
two dimensional scenario (thin films) or one dimensional scenario (thin
wires) then we are near the Compton wavelength and
one could expect neutrino like, bosonic behaviour, which is temperature
independent. Indeed it is known that the Dirac equation in two or one
dimensions, does indeed exhibit such features as massslessness and
helicity. Interestingly one can
expect similar behaviour at very low temperatures, because in this case
$\Delta p$ being negligible, $\Delta x$ is very large reminiscent of the
neutrino having a very large Compton wavelength and suggestive of the
superfluidity of Helium 3.
Another anomalous situation is when we have an assembly of nearly mono energetic
Fermions, in which case also $\Delta p \approx 0$. These
situations have been described elsewhere:\\
3. The Planck length, $10^{-33}cms$ shows up as the
gravitational radius of a Schwarzchild Black Hole, with mass $10^{-5}gms$.
Here general
relativistic phenomena which are purely classical meet purely quantum
mechanical phenomena:
For QMBH the radius is given by the Compton wavelength.
Let us first consider the equation,
\begin{equation}
\frac{Gm}{c^2} \sim \frac{\hbar}{mc}\label{e15}
\end{equation}
For the mass $m \sim 10^{-5}gms,$ (\ref{e15}) is consistent. It shows that at
the Planck scale the purely classical Schwarzchild radius equals the Quantum
Mechanical Compton wavelength, which as pointed out earlier, is at the root of
electromagnetism. This is borne out by the fact that at these mass and
length scales, we have, instead of (\ref{e3}),
$$\frac{Gm^2}{e^2} \sim 1$$
Ofcourse such particles with
life times $\sim \frac{\hbar}{mc^2} \sim 10^{-42}secs$ are at best virtual
particles which provide the underpinning for the foam like space of
geometrodynamics (cf.ref.2).
Alternatively, one could think of these particles as fluctuations of $\sim
N^{1/4}$, of the fluctuation of $\sim N^{1/2}$ in the number of particles.\\
4. We now start with the pre-space-time background
field of the instantaneous fluid or Ganeshas. From here we get the probability for
$N$ of them to appear as $\alpha \quad exp \quad [-\mu^2 N^2]$
This immediately ties up with the considerations of ref.5.,
if we identify $N$ with $x$. The justification
for this can be seen from the fact that the probability is non-negligible if
$$\Delta N \sim \frac{1}{\mu} \approx \frac{\hbar}{mc},$$
the Compton wavelength. Thus once again we conclude
that a probabilistic fluctuational collection of instantaneous particlets
or "Ganeshas" from a pre-space-time background shows up as a particle in space-time:
Probabilistic fluctuations lead to space-time and physics from
pre-space-time. Interestingly  the
Compton wavelength is a cut off at which the above Gaussian integral
to this length balances, the integral from here the
to infinity.\\
5. We now observe that the Bohm formulation discussed in detail in reference 1
converges to Nelson's stochastic formulation$^{10}$ in
the context of the QMBH. Indeed Bohm's non local potential as also Nelson's
three conditions merely describe the QMBH as a vortex, the mass being given
by the self interaction, the radius of the vortex being the Compton wavelength.
(cf.ref.1). We can get a clue to the origin of Quantum Mechanical fluctuations: Following Smolin we
observe that the non local stochastic theory becomes the classical local
theory in the thermodynamic limit, in which $N$ the number of particles
in the universe becomes infinitely large. However if $N$ is finite but
large, these fluctuations are of the order $1/\sqrt{N}$ of the dimensions
of the system, the universe in this case. Indeed this is exactly so because,
of equation (\ref{e9}). This provides a holistic rationale for the "spooky"
non-locality of Quantum Theory.\\
6. It was shown in ref.$^1$ that at the Compton wavelength, instead of the
electromagnetic (Kerr-Newman) field we get a QCD type potential,
$$ - \frac{\alpha}{r} + \beta r$$
with the correct ratio $\alpha/\beta$. As we have seen in this case we
encounter two or one dimensions. So instead of the three stress momentum
tensors $T^{\imath \imath}$ (each of which equals $\frac{1}{3} T^{00}$), which
leads to the electrostatic potential, we have only two or one of them so that
the charge will show up as $\frac{2}{3} e$ or $\frac{1}{3} e$, as in the case
of quarks. From the above potential we can also recover the quark mass. It
automatically follows that free quarks cannot be observed as we are dealing
with confined systems.\\
Considering now a large Compton wavelength (which as shown leads to the
neutrino's handedness), we have from the above potential,
$$g^2 \sim 6m c^2 \int T^{00} d^3 x',$$
whence
$$G \equiv g^2/m^2 \sim 10^{43} gm^{-2}$$
Thus in addition the correct coupling constant of the weak interaction along
with the additional feature of handedness emerges.\\
7. We could argue that the quantization of space-time which very roughly
leads to quantized wavelengths or frequencies is more fundamental than,
quantization of energy. Indeed this can be shown to lead to Planck's law.\\
8. In the spirit of point 5 above, if we use (\ref{e3}) in the form given before
(\ref{e9}) and (\ref{e9}) itself, we get,
$$R = \frac{e^2}{mc^2} \sqrt{N} = l \sqrt{N},$$
whence we can \underline{deduce} that,
$$l = \frac{e^2}{mc^2},$$
which is correct. This shows that the microphysical constants are not all
independent, e.g., the value of the Planck constant depends on $e,m \mbox{and}
c$. We could now argue that $'e'$ is given by the spinorial tensor density
$(N = 1$ of section 1) and further that $m$ follows in the QMBH model via, e.g.,
the pion Compton wavelength and classical electron radius equivalence (ref.1).
Then, a maximal universal velocity is the only requirement for characterizing
the entire universe, from the formation of electrons upwards.

\end{document}